\documentclass{pasj01}
\usepackage{mathrsfs} 
\usepackage{graphicx} 
\usepackage{mediabb}
\usepackage{color}  
  
\title{Molecular Bow Shock in the 3-kpc Norma Arm}  
\author{Yoshiaki \textsc{Sofue}\altaffilmark{} }
\altaffiltext{}{Institute of Astronomy, The University of Tokyo, Mitaka, Tokyo 181-0015}


\begin{document} 
\date{ } 
\maketitle   

\def\revise{}

\def\vlsr{v_{\rm LSR}} \def\v{v_{\rm LSR}} \def\Msun{M_\odot} \def\deg{^\circ} \def\r{\bibitem[]{}}     \def\/{\over}\def\kms{km s$^{-1}$}  \def\Vsun{V_0}  \def\Vrot{V_{\rm rot}}   \def\Tc{T_{\rm C}} \def\Tb{T_{\rm B}} \def\sin{{\rm sin}\ } \def\cos{{\rm cos}\ } \def\Hcc{ H cm$^{-3}$ }  \def\co{$^{12}$CO$(J=1-0)$ }\def\co{$^{12}$CO$(J=1-0)$ } \def\htwo{H$_2$} \def\sub{\subsection}\def\be{\begin{equation}} \def\ee{\end{equation}}\def\Kkms{K \kms} \def\Htwosqcm{H${_2}$ cm$^{-2}$}\def\Icut{I_{\rm c}} \def\Ico{I_{\rm CO}}\def\Icri{I_{\rm c}} \def\Ic{I_{\rm c}}
\def\Xco{X_{\rm CO}} \def\X{X_{\rm mass}} \def\XHI{X_{\rm HI}} \def\TbHI{T_{\rm b: HI}} \def\IHI{I_{\rm HI}}\def\Te{T_{\rm e}}\def\ne{n_{\rm e}}\def\ue{u_{\rm e}} \def\um{u_{\rm mol}}\def\ergcc{erg cm$^{-3}$}\def\nHH{n_{\rm H_2}}\def\mH{m_{\rm H}}
\def\sigv{\sigma_v} \def\x{\times}\def\Hcc{H cm$^{-3}$} \def\fmol{f_{\rm mol}}
\def\tauco{\tau_{\rm CO}}\def\tauhi{\tau_{\rm HI}}\def\Ts{T_{\rm S}}\def\Tco{T_{\rm B: CO}}\def\THI{T_{\rm B: HI}}\def\nH2{n_{\rm H2}}\def\nHI{n_{\rm HI}}\def\Halpha{ H$\alpha$ }\def\nuv{N_{\rm UV}}\def\ne{n_{\rm e}}\def\ni{n_{\rm i}}\def\nh{n_{\rm H}}\def\ar{\alpha_{\rm r}}\def\cs{c_{\rm s}}\def\Luv{L_{\rm uv}}\def\RS{R_{\rm S}}\def\rhii{R_{\rm HII}}\def\Lsun{L_\odot}\def\subsub{\subsubsection}
\def\fmol{f_{\rm mol}} \def\noi{\noindent}
\def\Htwo{H$_2$ } \def\sigv{\sigma_v } 
\def\rmkms{ \ {\rm km\ s^{-1}}}
\def\be{\begin{equation}} \def\ee{\end{equation}}
\def\Tn{T_{\rm n} } \def\Te{T_{\rm e} } \def\Ha{H$\alpha$ }
\def\({\left(} \def\){\right)} \def\[{\left[} \def\]{\right]}    
\def\Kkms{ K \kms } \def\Hsqcm{ H cm$^{-2}$ }
\def\Ico{I_{\rm CO} }
   
\def\tc{t_{\rm col}} \def\rhoc{\rho_{\rm cloud} } \def\tb{ t_{\rm bow} }
\def\rhom{\rho_{\rm mean} } \def\Rc{R_{\rm cloud} } 
\def\V{ V_{\rm rot} } \def\Vp{ V_{\rm p} } \def\Op{ \Omega_{\rm p} }
\def\Rbow{R_{\rm bow} } \def\Rhii{R_{\rm HII}}    \def\Rzero{R_0}    
\def\Lcone{L_{\rm cone} } \def\Acone{\Theta } \def\Lhii{L_{\rm HII}}
\def\K{{\rm K}} \def\pc{{\rm pc}} \def\hcc{{\rm H \ cm^{-3}}}

\begin{abstract} A molecular bow shock (MBS) at G24.4+00+112 $(l\sim 24\deg.4,\ b\sim 0\deg, \ \vlsr \sim 112$ \kms) is studied using the \co-line survey obtained with the Nobeyama 45-m telescope at $20''$ (0.71 pc) resolution. The terminal velocity uniquely locates the object at the tangent point of the 3-kpc expanding arm (Norma arm) with the distance of 7.3 kpc. The bow ridge extends over $\sim 160$ pc ($1\deg.3$) perpendicularly to the galactic plane, and is concave to a ring of HII-regions centered on G24.6+00 at the same distance. The edge on the down-stream (higher longitude) side of the MBS is extremely sharp, and is associated with several elephant trunks in gear-to-gear touch with the HII regions. On the up-stream (lower longitude) side of MBS, a broad HI bow is associated at the same velocity. The coherently ordered structure of HI, CO and HII gases indicates HI-to-\htwo transition at the galactic shock followed by efficient star formation due to dual compression, where the molecular gas is shock-compressed from up-stream side by galactic shock and from down-stream side by HII expansion. We propose a scenario of galactic sequential star formation (GSSF) along the spiral arms. We also discuss related ISM phenomena such as the hydraulic jump, bow shock, and Raileigh-Taylor instability occurring around the MBS. 
\end{abstract}    
 
\section{Introduction}

Galactic shock waves play the essential role in compressing the interstellar gas and triggering star formation (SF) in spiral galaxies (Fujimoto 1969; Roberts 1969). On the other hand, the current SF studies have been obtained either about individual molecular clouds (MC) and/or HII regions (Motte et al. 2018), or correlation analyses of the SF rate and gas density (Kennicutt and Evans 2012). As to the SF triggering mechanisms, the sequential SF in a molecular cloud (Elmegreen and Lada 1977) and cloud-cloud collisions (Scoville et al. 1986) appear to be considered most effective. This paper aims at linking the large- and small-scale SF mechanisms by studying the SF complex toward the tangent directions of the inner spiral arms in the Milky Way.

The first evidence for such a place that explicitly revealed galactic shock compression leading to star formation was obtained as the molecular bow shock (MBS) at G30.5+00+90 (after its galactic coordinates and radial velocity:  $l\deg,b\deg,\vlsr$ \kms; abbreviated G31) associated with the HII region W43 (Sofue 1985; Sofue et al. 2019), using the FUGIN CO-line survey (FOREST unbiased galactic ISM survey with the Nobeyama 45-m telescope: Umemoto et al. 2017). Also, numerous similar MBSs associated with giant cometary HII regions (GCH) have been found along galactic shock waves in the barred spiral galaxy M83 based on optical images taken with the Hubble Space Telescope (Sofue 2018).

In this paper we report the discovery of a new MBS G24.4 concave to the HII ring G24.6 in the tangential direction of the Norma Arm (3-kpc expanding ring). We name it MBS G24.4+00+112 (abbreviated as G24.4) after its galactic position and radial velocity, $l\sim 24\deg.4, b\sim 0\deg, \vlsr \sim 112$ \kms.

The HII ring G24.6 has been noticed as an association of bright HII regions centered on G24.6+00 with a diameter of $\sim 0\deg.5$ associated with CO and HI arc at the same velocity  (Sofue et al. 1984; Handa et al. 1986). The HII regions are further embedded in a diffuse shell-like HII region of diameter $\sim 0\deg.6$. Recocmbination-line velocities of the HII regions around G24.6 are measured to be $\vlsr = 96 \sim 117$ \kms (Downes et al. 1980), which locate the HII regions at the tangent point of galactic rotation at the same place as the MBS. Some of the HII regions and molecular clouds have been investigated individually using the FUGIN CO data in order to argue for cloud collision mechanism (Torii et al. 2018).

The radial velocities close to the terminal velocity make the distance determination unique without near/far ambiguity. The distance from the Sun of MBS G24.4+00+112 and HII ring G24.6 is estimated to be $d=R_0\cos \ l =7.28$ kpc for $l=24\deg.6$ and galacto-centric distance $r=R_0 \sin \ l=3.32$ kpc, adopting $R_0=8.0$ kpc for the galacto-centric distance of the Sun. The diameters of the HII ring and diffuse HII shell are $\sim 64$ ($0\deg.5$) and $\sim 76$ pc ($0\deg.6$), respectively.

We analyze the distribution and kinematics of molecular gas around MBS G24.4 using the FUGIN \co-line survey at a resolution of $20''$ on the used maps (0.71 pc at 7.3 kpc), as well as other CO surveys, HI, continuum and IR archival data. Based on the analyses, we discuss the HI-to-\Htwo transition, formation mechanism of the bow structure, and implication for the SF study. We propose a galactic sequential star formation (GSSF) scenario, with which coherent and long-lasting SF activity in the spiral arms of the Galaxy can be maintained.

\section{Data and Results}

\subsection{Norma arm on the longitude-velocity diagrams}

\begin{figure} 
\begin{center}        
\includegraphics[width=8.5cm]{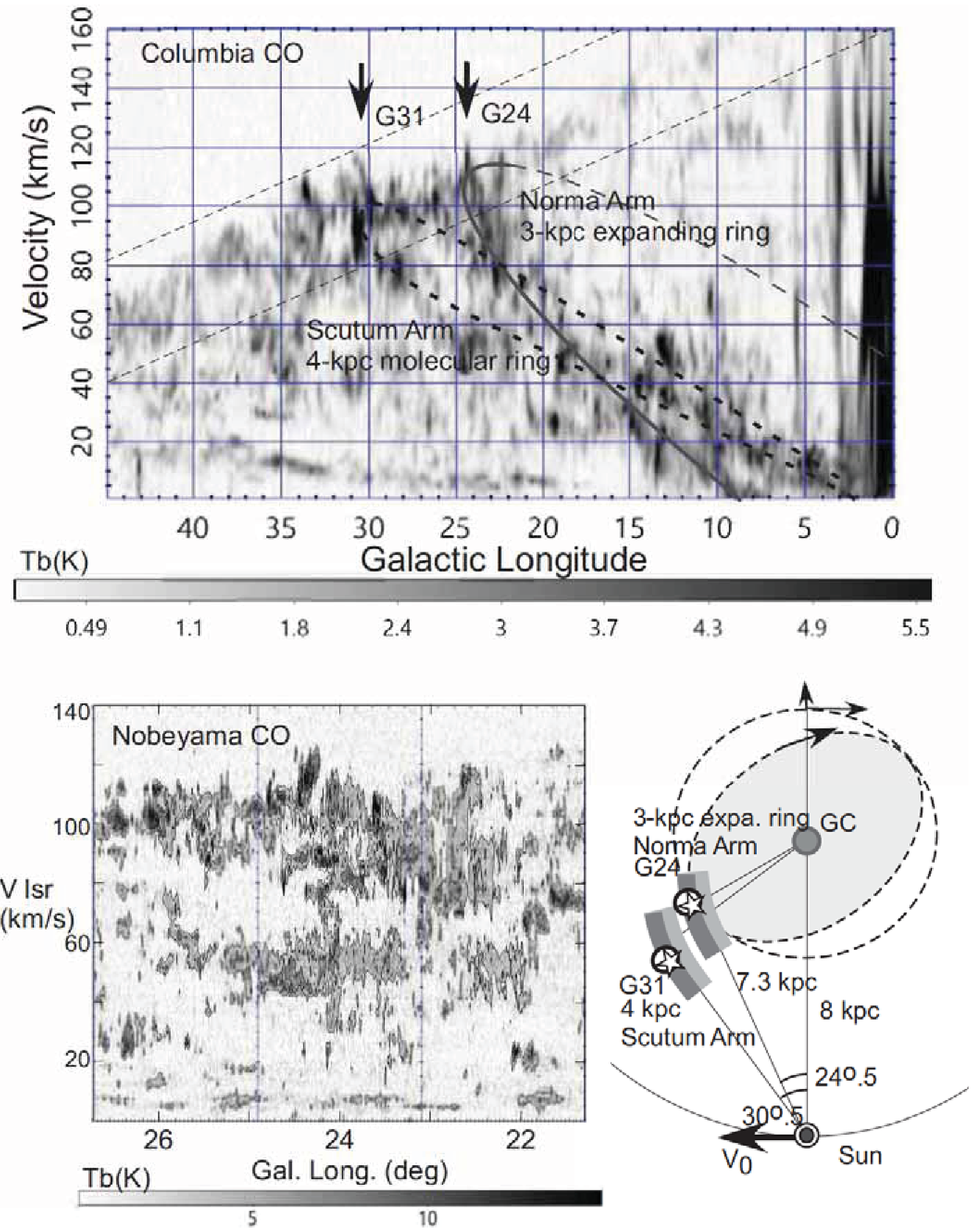}  
\end{center}
\caption{[Top] Longitude-velocity diagrams of \co $\Tb $ from Columbia 1.2m CO survey. G24+00 complex is seen as the bright CO cloud at ($l\sim 24\deg, \vlsr\sim 112$ \kms) at the tangent point of the Norma arm (3-kpc expanding arm/ring). 
[Bottom left] Longitude-velocity diagrams of \co $\Tb $ from Nobeyama 45m FUGIN CO survey. Note that the G24 cloud is significantly displaced from the mean terminal velocity of the surrounding gas.
[Bottom right] Illustration explaining the face-on locations of the Norma and Scutum arms. Dashed ellipse and circle with velocity arrows represent oval orbit in a bar potential and expanding ring, respectively.}
\label{fig1}   
	\end{figure}

We start with showing a longitude-velocity (LV) diagram of the first quadrant, revealing the two densest spiral arms in the inner Galaxy. Figure \ref{fig1} shows a \co-line LV diagram from the Columbia survey (Dame et al. 2001), and figure \ref{fig1} is a close up of the G24 region from FUGIN CO survey. The LV diagrams reveal two most prominent arms, one is the Scutum arm, and the second brightest arm is the Norma arm, which is the subject of this paper, as traced by the inserted lines.
        
The Norma Arm, also known as the 3-kpc expanding arm/ring, is recognized as a bright elliptical ridge running from $(l,v)\sim(0\deg,-50 \rmkms)$ through $(24\deg,110 \rmkms)$. The G24.6+00 HII ring is located at high-velocity end of this LV ridge. Coinciding with this position and velocity, we find a bright knot of CO emission at $(l,v)\sim 24\deg.4, 110-122 \rmkms)$. 

The terminal-velocity region is enlarged in the FUGIN LV diagram, where the CO brightness is as high as $\Tb \sim 15$ K at $(24\deg.4, 115 \rmkms)$. The radial velocity of the CO clump exceeds significantly the mean terminal velocity of the surrounding CO emission, indicating non-circular motion and/or a local velocity anomaly.

\subsection{Tangent-point CO Maps}

\def\vtan{v_{\rm tan}}

We then make distribution map on the sky of molecular gas having the terminal velocities of galactic rotation, which represents the cross section of the galactic disk along the terminal velocity circle without suffering from the near- or far-side ambiguity. The "tangent-point" map uniquely reveals the gas distribution as a function of the galacto-centric distance $R$ and height from the galactic plane $z$ through $R=R_0 \sin l$ and $z=r \cos b$, where $R_0=8.0$ kpc is the distance of the Sun from the Galactic Center (GC).

	\begin{figure*} 
\begin{center}         
\includegraphics[width=13cm]{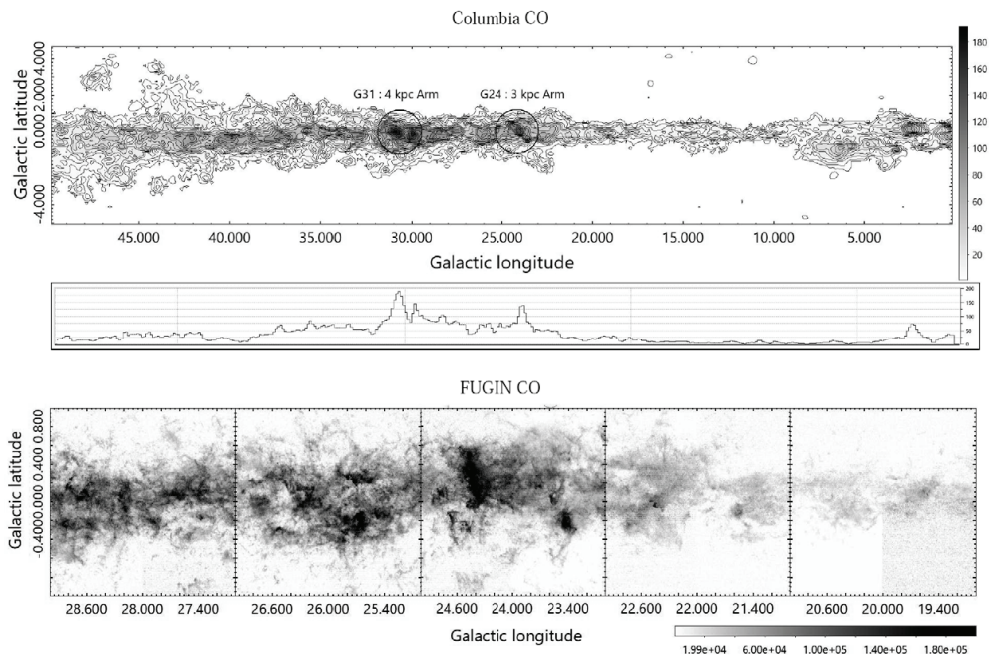} 
\end{center}
\caption{[Top] Columbia 1.2m $\Ico$ tangent-point map integrated between the two dashed lines in figure \ref{fig1} ($\sim \pm 20$ \kms of $\vtan$, showing a global distribution of the terminal-velocity molecular gas. The bottom panel shows intensity variation along the galactic plane. Note the two bright molecular peaks at G31 and G24, representing the cross sections of the 4-kpc molecular ring (Scutum arm) and the 3-kpc arm (Norma arm), respectively.
[Bottom] Nobeyama 45m FUGIN $\Ico$ tangent-point map around G24 integrated between $\pm 15$ \kms about $\vtan$. G24 molecular complex is outstanding among the neighboring clouds and disk. Grey scale is in K m s$^{-1}$}
\label{fig2} 
\label{fum0}   
	\end{figure*}   

Figure \ref{fig2} shows a thus obtained tangent-point map, where is shown the \co line intensity $\Ico$ integrated between $\vtan-20$ to $\vtan+20$ \kms as obtained from the Columbia CO survey (Dame et al. 2008), where $\vtan$ is the terminal velocity at $l$ from smoothed rotation curve. The bottom panel of the figure shows variation of the CO intensity along the galactic plane. 
Figure \ref{fig2} is a close up of the G24 region by a tangent-point $\Ico$ map of the \co line integrated within $\vtan-15$ to $\vtan+15$ \kms obtained using the FUGIN CO survey. The map reveals a dense arc-shaped ridge at G24.4, extending over $1\deg.5$ in the latitudinal direction.

These tangent-point maps reveal two prominent peaks at G31 ($l\sim 30\deg.5$) and G24.4 ($\sim 24\deg.4$), which are the most prominent and densest molecular regions in the inner Galaxy, representing the tangential directions of the two major spiral arms in the first quadrant of the Galaxy.
G31 corresponds to the tangential direction of the 4-kpc molecular ring, or the Scutum Arm, which has recently been extensively analyzed in view of the galactic shock waves and molecular bow shock (Sofue et al. 2019). 
G24.4 is the second-brightest molecular region toward 3-kpc expanding ring/arm, or the Norma Arm, and is the subject in this paper.

Another prominent feature observed in this figure is the vacant region of the CO  emission at $l<\sim 22\deg$, exhibiting a molecular-gas vacant region inside $R\sim 3$ kpc from the Galactic Center (GC). This vacancy has been studied and interpreted as the "3-kpc galactic crator" produced by an explosive mass sweeping by the GC activity (Sofue 2017).

\subsection{Channel maps}
        
The distribution of CO brightness near the terminal velocity can be obtained by sliced maps (channel maps) at different velocities. In figure \ref{figA}, we show channel maps of a $2\deg \times 2\deg$ region around G25+00 at various velocities. The G24.4 CO clump is now resolved in velocity, showing up as a vertically (in the latitude direction) extended bright ridge at $\vlsr = 105 - 118$ \kms with the most prominent appearance at 112 \kms. 

\begin{figure*} 
\begin{center}       
\includegraphics[width=13cm]{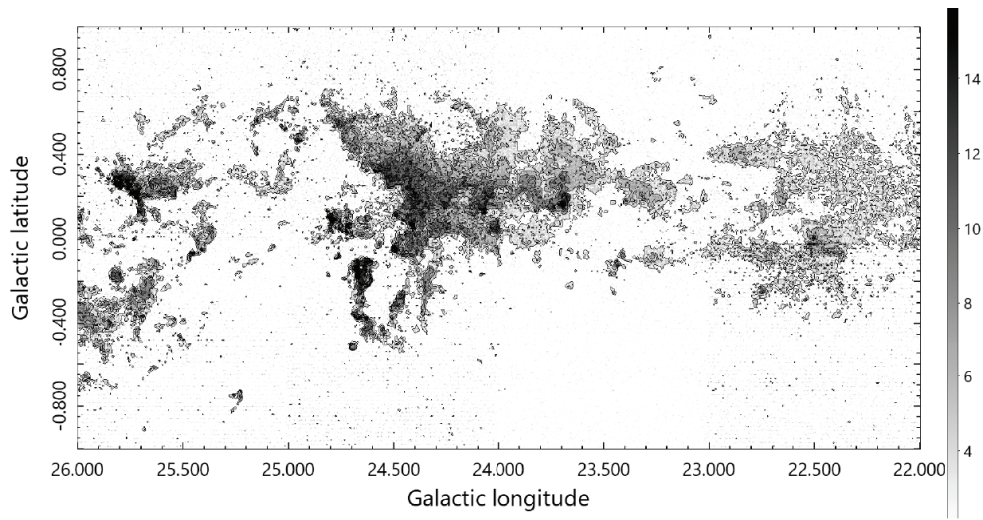}    
\end{center}
\caption{\co $\Tb $ map at $\vlsr =112$ \kms at the tangent point of the Norma arm, also known as the 3-kpc expanding ring, showing the molecular cross section. The arm shows up as the large CO line clump at $\sim$G24 with a sharp bow-shaped eastern edge indicative of a shock front. This bow structure is called the molecular bow shock (MBS) G24.4+00+112. Contours are every 2.5 K in $\Tb$.}
\label{fig3} 
	\end{figure*}

A channel at $\vlsr=112$ \kms is enlarged for wider area in figure \ref{fig3}.  The G24.4 CO clump is now resolved to show a sharp arc-shaped structure concave to the center of the G24.6+00 HII ring. The CO emission extends from $b=-0\deg.5$ to $+0\deg.8$, making an arc (bow) of length of $\sim 1\deg.3$ ($\sim 160$ pc) perpendicular to the galactic plane. 

	\begin{figure} 
\begin{center}       
\includegraphics[width=8.5cm]{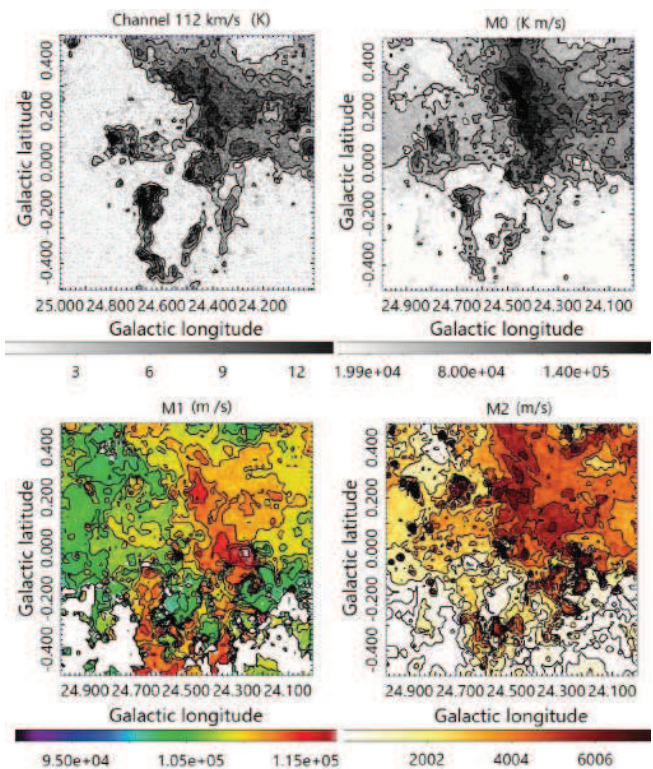}   
\end{center}
\caption{$\Tb$ map at 112 \kms (top left), intensity integrated at $100\le \vlsr \le 126$ \kms (moment 0 map, top right), velocity field in m/s (moment 1, bottom left), and velocity dispersion in m/s (moment 2, bottom right) .}
\label{fig4} 
	\end{figure}

Figure \ref{fig4} shows the channel map at 112 \kms compared with moment 0, 1 and 2 maps around G24.5 for 1 square degree region, showing the CO-line intensity integrated from $\vlsr=100$ \kms to 126 \kms, radial velocity field, and velocity dispersion in the same velocity range, respectively.

The eastern edge (inner edge) of the arc at G24.4 is extremely sharp, and is associated with elephant-trunk features avoiding HII radio peaks. 
Figure \ref{fig7} shows a horizontal cross section of the CO brightness at 112 \kms at $b=0\deg.176$. The cross section shows an extremely sharp cut-off of the intensity toward the east.

The sharp brightness peak at the tangent point of the spiral arm is just inverse to what expected for a usual galactic shock wave caused by a supersonic inflow from the up-stream side (from right, or from lower longitude side). This particularly sharp edge in the reverse side of the galactic shock may imply an additional compression from the down-stream side (from the left) likely induced by an expansion of HII gas from the star forming sites in the G24 HII ring.

The western side of the CO edge extends rather broadly, making a large bow-shaped ridge, and extends for about $1\deg$ till $l\sim 23\deg.5$ with the vertical extent decreasing. We call this structure the G24.4 MBS (molecular bow shock). As a whole, G24.4 MBS looks like a bow shock caused by an inflow from the west, being stopped by an expanding spherical front from the east around the G24.6 HII ring.

\subsection{HI map and molecular fraction}

\begin{figure}  
\begin{center}      
\includegraphics[width=8cm]{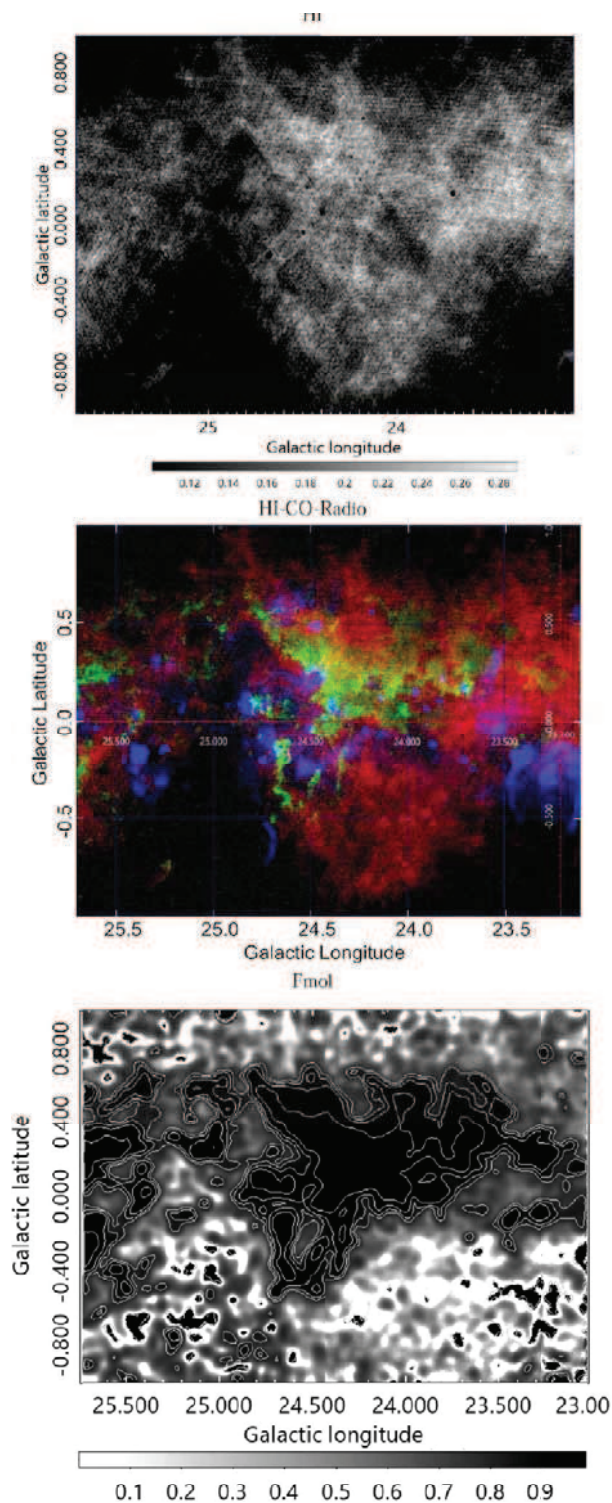}  
\end{center}
\caption{[Top] HI brightness at 112 \kms from the THOR survey. Intensity unit is the brightness in mJy/beam. 
[Middle] Composite color map of HI (THOR) and CO (FUGIN) brightness at 112 \kms in red and green, respectively, and 1.4 GHz continuum (VGPS) in blue. Continuum intensity is non-linearly enhanced for stronger sources. Brightness is adjusted appropriately to show the mutual spatial relations of the emissions. 
[Bottom] Molecular fraction $\fmol$ map. Contours are drawn from 0.8 to 0.95 at 0.05 interval. The MBS is almost molecular with $\fmol \ge 0.95$. } 
\label{fig5} 
\end{figure}

For the high velocity resolution and comparable angular resolution, the HI data from the HI/OH/Recombination line survey of the Milky Way (THOR) (Beuther et al. 2016 ) can be directly compared with the present CO maps. Figure \ref{fig5} shows a brightness map of the HI line emission at $\vlsr = 112$ \kms, and figure \ref{fig5} shows an overlay of HI in red, \co brightness in green at the same velocity (same as figure \ref{fig3}), and a radio continuum map at 1.4 GHz in blue color from the VLA Galactic Plane Survey (VGPS: Stil et al. 2004). 

The HI map shows a cavity centered on G25+00, around which a broad, arc-shaped concave clump is distributed  from $l=23\deg.6$ to $25\deg.6$, where the vertical extent gets significantly lager than that at lower longitudes (up-stream side), forming a broad peninsula toward negative latitude. The eastern edge concave to G25 of the HI peninsula appears to trace approximately the G24 MBS arc, while the CO arc is much sharper.

The galactic shock theory predicts that the ISM in-flowing from the up-stream side is compressed near the galactic shock front and forms a broad bow structure (Martos and Cox 1998). As the gas is condensed, HI is transformed to molecules, increasing the molecular fraction. This is indeed observed around G24 in the present data, where the HI gas flows from the lower longitude (right) side, encounters the 3 kpc arm to form the HI arm, and is transformed to \Htwo to form the MBS at G24.

We have calculated the molecular fraction $\fmol$ from the brightness of CO and HI at 112 \kms using the following equation.
\begin{equation}
\fmol={2n_{\rm H_2} \over n_{\rm HI}+2n_{\rm H_2}}
={2 \Xco \Tb ({\rm CO}) \over X_{\rm HI} \Tb({\rm HI}) +2 \Xco \Tb({\rm CO})},
\end{equation}
where $\Xco=2\times 10^{20}\ {\rm cm^{-2} [K\ km\ s^{-1}]^{-1}}$ is the CO intensity-to-H$^2$ column density conversion factor, $X_{\rm HI}=1.82 \times 10^{18} {\rm cm^{-2} [K\ km\ s^{-1}]^{-1}}$ is HI intensity to column density conversion factor. The HI brightness temperature $\Tb(\rm HI)$ was estimated from the HI brightness in Jy per $40''$ beam used in the THOR archive. Calculated result is shown in figure \ref{fig5}, which indicates that the molecular fraction in the CO bright region is as high as $\fmol \ge 0.95$, so that the ISM is almost totally molecular inside the galactic shock wave.

\subsection{Radio continuum and IR maps for HII regions}

The radio continuum loop of HII regions, G24.6+00, which was called the "Scutum ring" of HII regions in our earlier paper (Handa et al. 1986), is composed of five bright radio sources embedded in a round. shell-like extended component. The sources, including the extended emission, have flat radio spectra between 2.7 and 10 GHz, indicating that they are HII regions.

Recombination-line observations of the radio sources have shown that their radial velocities lie between 97 to 117 \kms, showing that they are at the same kinematical distances to the presently observed molecular and HI bow features. 
The five sources are distributed along the maximum ridge of the extended shell component. We hereafter call the entire structure, including the compact HII regions and HII shell, the "G24.6+00 HII shell".

The extended component has a radial brightness distribution typical for a spherical shell of outer radius $\sim 0\deg.3$ (38 pc) as measured at the half-maximum intensity point at 5 and 10 GHz (Handa et al. 1986). The shape and cross section are remarkably similar to those of the Rosette nebula (e.g., Celnik 1986), except that the radius of G24 (38 pc) is 2.2 times that of Rosette (17 pc) and hence the volume and energy are $2.2^3 \sim 11$ times. Also, the G24 shell is superimposed by the compact HII regions.

\begin{figure*} 
\begin{center}   
\includegraphics[width=11cm]{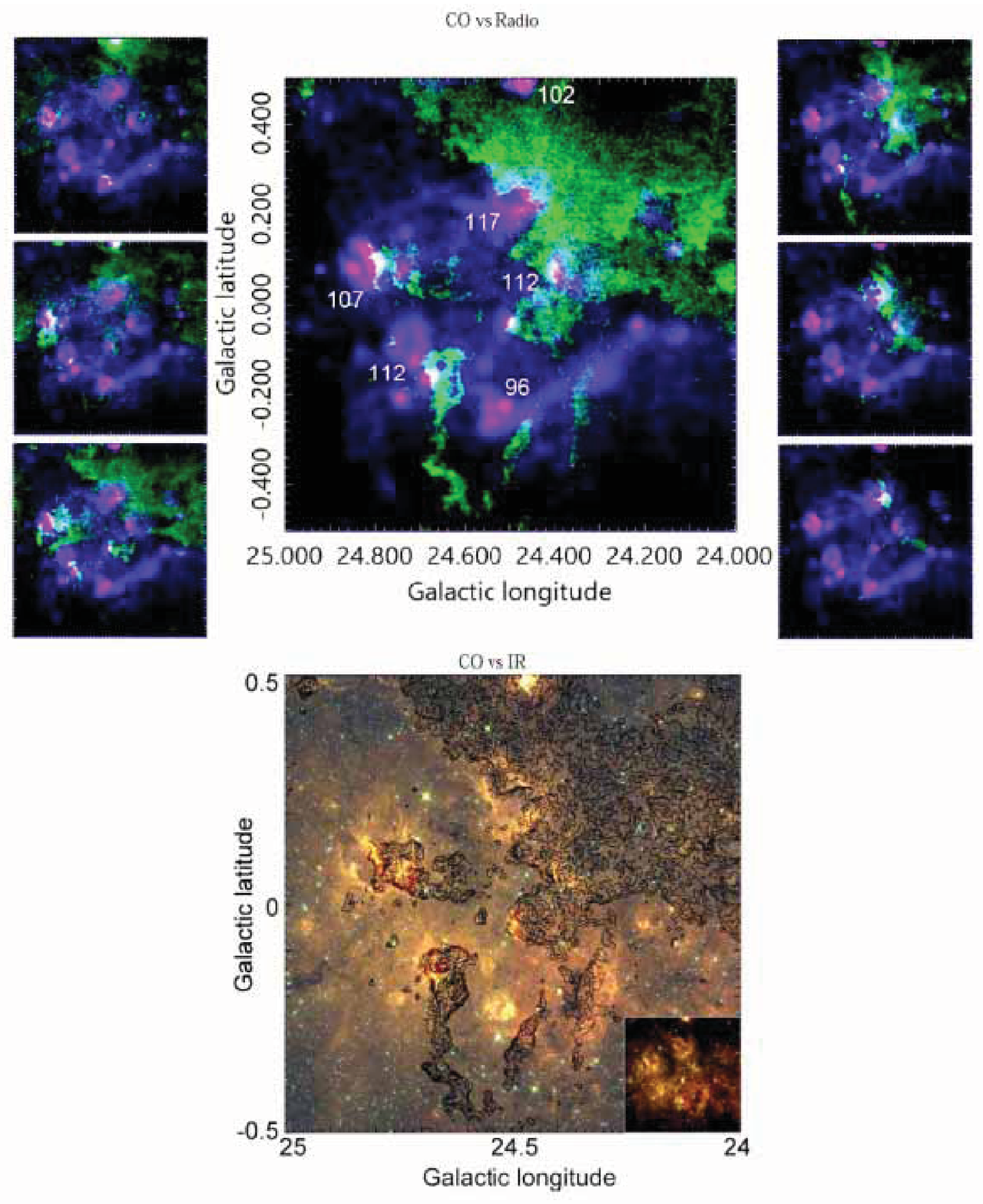}  
\end{center}
\caption{[Top center] Overlay of 1.4 GHz continuum map from VGPS on the \co channel maps at $\vlsr=112$ \kms, 102, 106,  109 (top left), and 115, 119, 122 \kms (right).  For continuum intensity scale, see \ref{fig10}.
[Bottom] Overlay of \co 112 \kms brightness contours (every 2 K from $\Tb=4$ K) on GLIMPSE 4.5, 5.8 and 8 $\mu$m composite map downloaded from https://third.ucllnl.org/cgi-bin/gpscutout. Note the CO "elephant trunks" with their high-density heads associated with compact HII regions on their down-stream (higher longitude) sides. Inserted small is a clearer image without contours to show the round shape of the diffuse HII region.  } 
\label{fig6}
\end{figure*}

We now compare the CO maps with the radio continuum map from the VLA Galactic Plane Survey at 1.4 GHz (VGPS: Stil et al. 2006 ) at an angular resolution of $25''$, comparable to that of FUGIN $(20'')$ in figure \ref{fig6}.

The MBS composes a large CO-bright bow on the up-stream (lower longitude) in touch with the western edge of the HII ring composed of compact continuum sources and round extended emission centered on G24.6+00. The MBS is concave to the continuum ring with the sharp inside edge facing the continuum edge. 

The inside edge of MBS at G24.4 shows wavy fluctuation associated with elephant-trunks extending toward inside the bow. Two major compact HII regions at G24.45+0.2 and G24.40+0.1 are located deep inside the "bays" between the trunks. 

Except for these two sources, most of the continuum sources (HII regions) are located on the down-stream (higher longitude) sides of individual molecular clouds and/or trunk heads closely in touch from the eastern sides. Also remarkable is that the eastern edges of the molecular clouds/tongues are sharper than west, and are often concave to their associated HII regions.         
         
The spatial relation between the CO gas and HII regions can be also confirmed by comparing with the hot dust emission in infrared images. Figure \ref{fig6} shows an overlay of the CO contours at 112 \kms on the pseudo color image of 4, 5 and 8 $\mu$m emissions as reproduced from the Spitzer/GLIMPSE survey (Galactic Legacy Infrared Mid-Plane Survey Extraordinaire: Churchwell et al. 2009). The IR features are similar to those of radio continuum, showing both the compact HII regions on a ring and the extended shell-like emission resembling the Rosette nebula.
        
\subsection{Kinematics and Energetics} 

\def\Mmbs{M_{\rm MBS}} 

\begin{figure} 
\begin{center}      
\includegraphics[width=6cm]{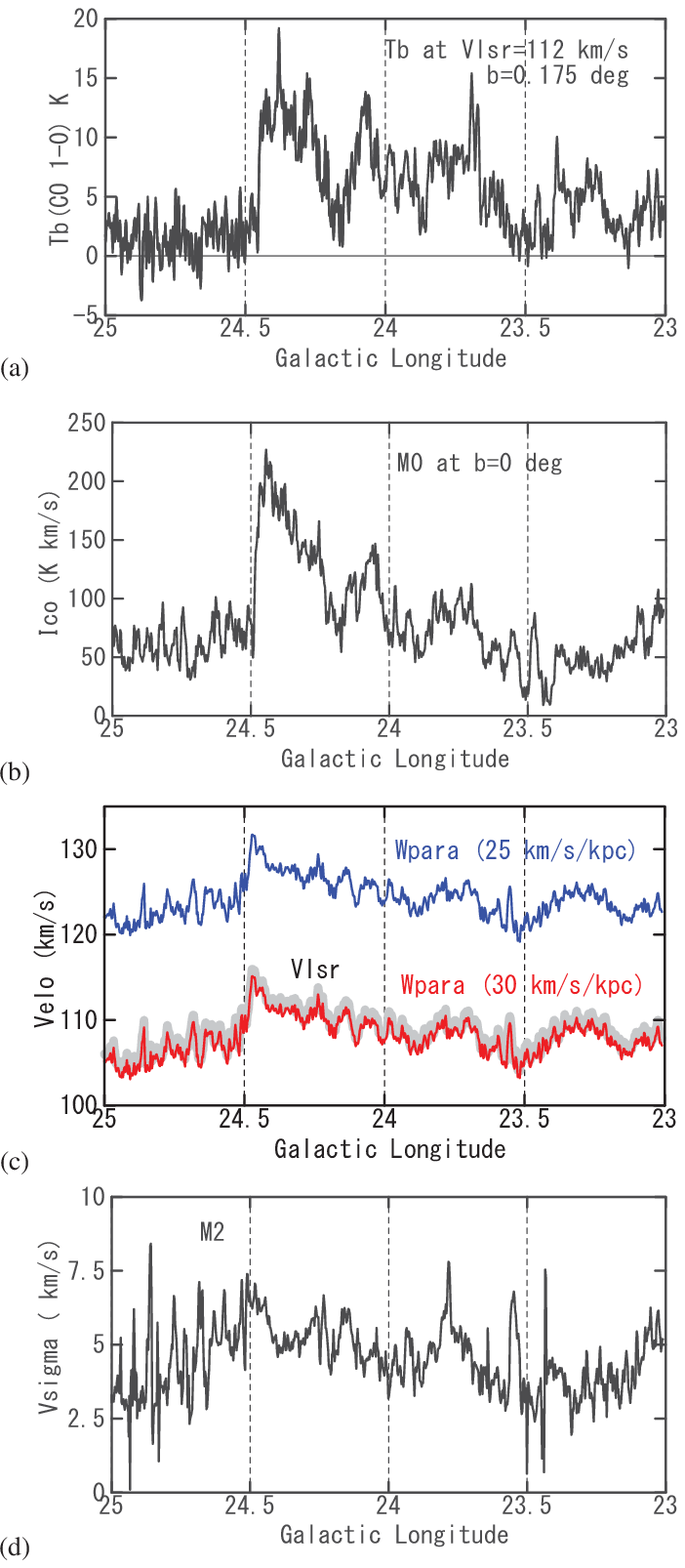} 
\end{center}
\caption{(a) Longitude variation of $\Tb$ of the \co line emission at 112 \kms, showing an extremely sharp eastern edge of the MBS at G24.6.  
(b) Longitudinal variations of $\Ico$ integrated from 100 to 126 \kms at $b=0\deg.185$, (c) $\vlsr$ and $W_{//}$ (red) for $V_0=238$ \kms and $\Omega_{\rm p}=30$ (red) and 25 (magenda) \kms kpc$^{-1}$, and (d) $v_\sigma$ (M2). }
\label{fig7}  
\end{figure}

Kinematical behavior of the MBS can be learned from the radial velocity distributions. Figure \ref{figA} shows LV maps at various latitudes. In figure \ref{fig4} we showed the maps of moment 1 (mean velocity) and 2 (velocity dispersion).
Figure \ref{fig7} shows cross sections along longitude at $b=0\deg.19$ for $\Ico$, $\vlsr$ and $v_\sigma$. 

The radial velocity and velocity dispersion attains their sharp maximum in coincidence with the sharp peak of the intensity at the inner edge of the MBS at $l=24\deg.5$ in figure \ref{fig7}. The steep jumps of the intensity and velocities are coherently observed along the inner arched edge of the MBS in figure \ref{fig4}, which indicates that the gas is shocked along the MBS's inner edge.

These velocity structures across the MBS can be understood as due to the galactic shock wave and/or the expansion of hot gas from the G24.6+00 HII complex. Below we discuss several possibilities to cause the observed kinematics of MBS.

The mass of the MBS can be estimated from its extension on the sky and CO intensity. We here approximately measure the extension as $1\deg \times 0\deg.2 \sim 130\ {\rm pc} \times 25$ pc at the unique distance of 7.3 kpc, averaged CO intensity of $\Ico \sim 150$ K \kms. Then, for a conversion factor of $\Xco=2.0 \times 10^{20}$ \Htwo cm$^{-2}$ [K \kms]$^{-1}$, we obtain a column density $N_{\rm H_2} \sim 3\times 10^{22}$ \Htwo cm$^{-2}$, volume density of hydrogen $n\sim 400$ \Hcc taking a line of sight depth of the shell of $\sim 50$ pc, and total mass of the MBS $M_{\rm MBS}\sim 1.55 \times 10^6 \Msun$. 
Since this amount of gas is expanding at $\sim 10$ \kms, its kinetic energy of expanding motion is estimated to be $E_{\rm exp}\sim 1/2 M_{\rm MBS} v_{\rm exp}^2 \sim 1.55\times 10^{51}$ erg.

A possible mechanism to form the molecular shell structure is compression of gas by an expanding HII region at sound velocity, $v_{\rm exp}\sim 10$ \kms. We assume that the radius of the expanding HII sphere $\Rhii$ is equal to that of the diffuse shell-like HII region, as identified by Handa et al. (1986). Using the radial brightness distributions at 10 and 5 GHz (Handa et al. 1986), we measured the half-intensity radius of diffuse HII shell to be $\sim 0\deg.3$ (38 pc).

We, then, obtain the age of the HII shell to be about $\tau_{\rm HII}\sim R_{\rm HII}/v_{\rm exp}\sim 4\times 10^6$ y for $\Rhii=38$ pc and $v_{\rm exp}=10$ \kms. This time scale is much longer than that due to SN explosion, and would be sufficient for star formation to proceed and HII regions to develop in the compressed molecular shell, as indeed observed as compact radio sources. This idea is consistent with the scenario proposed by Handa et al. (1987). 

In order for this mechanism to work for the present HII shell, the required luminosity of the central OB cluster is estimated to be $\Lhii \sim 1\times 10^4 \Lsun$, typical for a big HII region, assuming $\Tn\sim 20$ K, $\Te \sim 10^4$ K, and inserting $n\sim 400$ \Hcc and $\Rhii\sim 38$ pc (Sofue 2018, Sofue et al. 2019) 

The velocity jump at G24.6 of MBS observed in the LV diagram (figure \ref{fig1}) and moment 1 plot (figure \ref{fig7}) is difficult to explain by the galactic shock compression alone. The angular momentum conservation of gas flow requires
\be
VR=(V+\delta V)(R+\delta R),
\ee
where $\delta V\sim \delta \vlsr\sim \delta W_{//}$ and $\delta R$ are displacements of the rotation velocity and radius from the mean rotation (Roberts 1969). The galactic shock in a trailing arm predicts $\delta R>0$ so that $\delta V<0$, which is contrary to the observed positive anomaly of $\vlsr$. 

In order for $\delta V$ to be positive as observed, $\delta R$ must be negative. Since $\delta V\sim +5$ \kms, we may estimate that $\delta R \sim -R \delta V/V \sim -70$ pc for $V\sim 210$ \kms. This implies that the gas in MBS was accumulated "backward" from the eastern side of MBS, likely by expansion of the HII region G24.6. However, the estimated displacement (70 pc) is about twice the HII radius. This discrepancy may be eased, if a part of $\delta V$ includes the expansion velocity of HII region itself such that the HII center and MBS are not aligned on the sky perpendicular to the line of sight. Some tens of degrees would be sufficient to cause a few \kms to explain the additional component of $\delta V$.

\section{Discussion}

\subsection{Galactic shock theory}

\def\Vpat{V_{\rm pattern}}
\def\Wpara{W_{//}}

The galactic shock theory predicts that the azimuthal velocity of gas with respect to the pattern speed of the spiral arm  $W_{\rm //}=\Vrot - \Vpat$, where $\Vpat=r\Omega_{\rm p}$, varies coherently with the variation of the density as a function of the displacement from the potential minimum along the flow line (Roberts 1969). 

\begin{figure} 
\begin{center}        
\includegraphics[width=5.5cm]{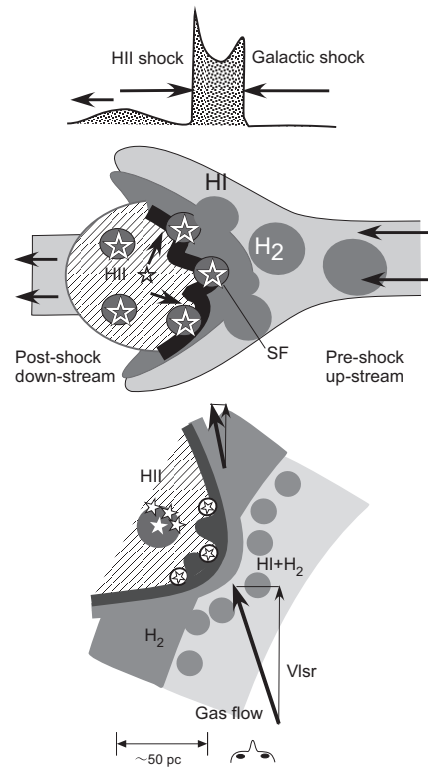}   
\end{center}
\caption{Schematic view of the MBS. The bottom panel shows a face-on view, showing the flow line. The radial velocity changes from red- to blue-shifted with respect to the shock front. This is indeed observed in the velocity field (moment 1 map) of the bottom-left panel of figure \ref{fig4}. The velocity dispersion increases at the shock, and decreases after the passage to become laminar flow, which is also observed in the moment 2 map in figure \ref{fig4}.  }
\label{fig8}
\end{figure}

Recalling that $V_{\rm rot}=\vtan+V_0\ \sin \ l$ and $r=R_0 \sin l$, and that measured $\vlsr$ is nearly equal to $\vtan$, $\vlsr \simeq \vtan$, we have 
 \be
 W_{//} = \Vrot - r \Omega_{\rm p}
 =\vlsr +(V_0-R_0 \Omega_{\rm p})\sin \l,
 \label{wpara}
 \ee
 where $\Omega_{\rm p}$ is the angular pattern speed of the spiral shock wave. 
 {\revise 
 Figure \ref{fig7}(c) shows thus calculated "observed" $\Wpara$ by adopting reasonable values for the pattern speeds of $\Omega_{\rm p}=25$ and 30 \kms kpc$^{-1}$. If the corotation radius of the Galaxy is equal to $R_0$ (e.g. Mishurov et al. 1997), we have $W_{//}\simeq \vlsr$, or $\vlsr$ is a direct measure of  $W_{//}$, which occurs when $\Omega_{\rm p}=238{\rm km\ s^{-1}}/8 {\rm kpc}\simeq 30$ \kms kpc$^{-1}$. Even if at different $\Omega_{\rm p}$ around this value, $\vlsr$ is an approximate measure of the flow.
  
Thus, figure \ref{fig7}(c) shows that the flow velocity with respect to the density wave potential increases from the up-stream (lower longitude) side toward down-stream, and it suddenly decreases after the passage of the shock front at G24.5. This behavior is in accordance with the variation of the CO brightness and intensity, and hence with the variation of gas density. 
Figure \ref{fig8} illustrates schematically the variation of velocity vectors across the shock front.

A the same time, the velocity dispersion (moment 2) increases toward the shock front, and decreases after the passage, indicating the change from turbulent to laminar flows (figures \ref{fig7}(d) and \ref{fig4}). 
Such an increase in the specific turbulent energy at the shock wave is indeed modeled by Mishurov (2006), and must play an additional role to trigger the star formation through cloud-cloud collisions.
}

This shock produces the global enhancement of the gas density, causing the large-scale bumps of CO and HI, already starting at $\sim$G23 (figure \ref{fig2}). This results in HI-to-\Htwo transition, as revealed by the $\fmol$ map in figure \ref{fig5}, and makes the highest $\fmol$ region near the MBS. 

As the compression and transition to \Htwo proceed, star formation may have taken place at $\sim$G24.6 to produce an intense HII region, as indicated by the extended radio continuum emission. As the star formation proceeds, the HII shell expands to push the surrounding molecular gas, and push back MBS to produce its concave structure with sharp inner edge. At the back-compressed front of the MBS on its down-stream edge, most recent star formation takes place, producing the compact HII regions in close touch with the MBS eastern edge in teeth-to-teeth contact. The shock interface between the HII gas and MBS is now observed as the sharp jumps of density and velocity at G24.6 shown in figure \ref{fig7}.

\begin{figure}
\begin{center}        
\includegraphics[width=7cm]{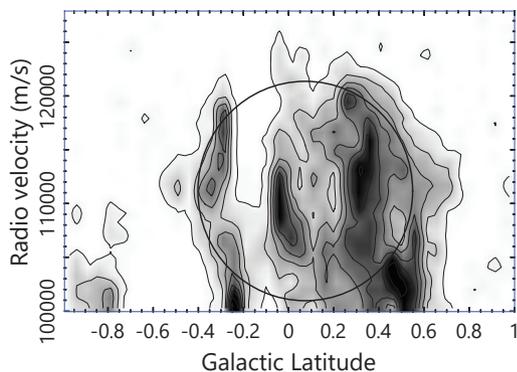}      
\end{center}
\caption{Latitude-velocity diagram averaged over 11 channels around $l=24\deg.5$ (intensity scale is arbitrary), showing a round ridge as traced by the circle, indicative of expansion of the MBS at $\sim 10$ \kms when cut across G24.6.}
\label{fig9}
\end{figure}

The expansion of the shock front in the MBS caused by the expanding HII gas may be observed as a cross section in a position-velocity diagram in the latitude direction. Figure \ref{figA} shows such $b$ vs $\vlsr$ (BV) diagrams at different longitudes, and figure \ref{fig9} is a BV diagram averaging 11 channel BV diagrams around G24.6. A round ridge of the CO emission is recognized as traced by the inserted circle centered on G24.6+0.1+112 with the diameter of $0\deg.8$ and 20 \kms. This may be interpreted as due to an expanding shell of radius $0\deg.4$ (50 pc) at a velocity $\sim 10$ \kms.

\subsection{Bow shock and hydraulic jump}

\def\RS{R_{\rm bow}}

{\revise 
Based on the coherent distributions of the HI, \Htwo, and HII gases in the G24 region as revealed in this paper, we may draw an image of the cross section of a galactic shock wave as follows. 

At the galactic shock front, the gas is not only compressed in the longitudinal direction, but also pushed away toward the halo by a hydraulic jump, which makes the vertically extended HI and molecular spurs, composing a large scale bow structure (Tosa 1973; Martos and Cox 1998; Mishurov 2006).

Besides the compression by the galactic shock, the gas is further shock-compressed and bent to make a sharper MBS, when the molecular spur interacts with pre-existing HII regions, which are expanding backward at supersonic velocity.
In our recent papers (Sofue et al. 2019; Sofue 2018) we applied the stellar bow shock theory by Wilkin (1996) to an MBS. The front curvature $\RS$ at the bow head, or the stand-off distance from the central engine, is related to the injection rate of gas, ambient density and inflow velocity of gas, which are further related to the observed MBS parameters.  
}

Figure \ref{fig11} shows the calculated bow fronts for some different $\RS$ values, where the bow shape satisfies self-similarity, so that the curves are identical except for the bow head curvature ($\RS$).  
The front appears to fit the observed MBS shape, if $\RS\sim 40$ pc. Using the formula in the above papers, such a radius can be realized by a parameter combination with a pitch angle of the arm $p\sim 12\deg$, inflow velocity of gas to the MBS $V_{\rm flow}\sim(V-V_{\rm p})\sin\ p \sim 20$ \kms for $V\sim 210$ \kms and $\Omega_{\rm p}\sim 30$ \kms kpc$^{-1}$, and gas density before shocked $\rho\sim 100$ \Htwo cm$^{-3}$, although the combination is not unique. 

 	\begin{figure} 
\begin{center}   
\includegraphics[width=7cm]{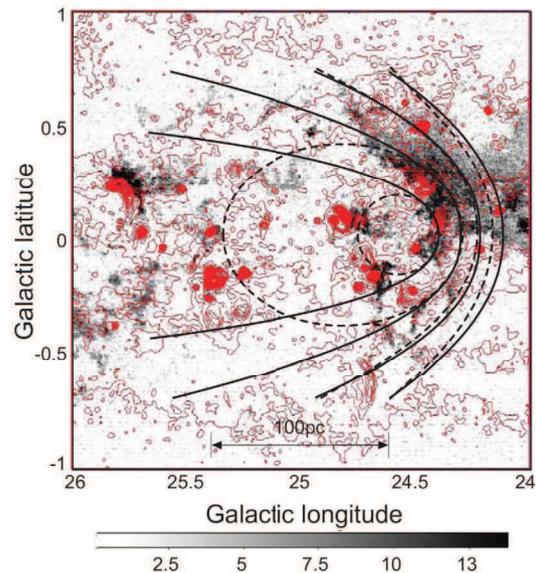}   
\end{center}
\caption{Model fronts of bow shock (full line) and HII region (dashed line) for different injection energies overlaid on the \co $\Tb$ map at 112 \kms in grey scale (K) and 1.4 GHz continuum map from VGPS by contours every 2 mJy/beam starting from 20 mJy/beam. }
\label{fig10}
\end{figure}

\subsection{Galactic Sequential SF (GSSF)}

As HII regions expand in the galactic shock, their up-stream side encounters the MBS and is decelerated, whereas the down-stream side expands more freely. The bow head of the MBS encountering the expanding HII front, thus, becomes a place of active star formation as known for a blistered HII region contacting a GMC (Elmegreen and Lada 1977). The MBS is further compressed from the up-stream side by the galactic shock wave. Thus, the MBS becomes a place where the interstellar gas is  shock-compressed from up- and down-stream sides. The dual compression, by galactic shock plus feedback from the born OB stars and HII regions, works coherently in the galactic spiral arms, and triggers the star formation at higher rate than the currently considered for cloud-collision mechanism and/or simple galactic shock. 

On individual HII region size, the SF rate has been considered to be determined by cloud collisions (e.g., McKee and Ostriker 2007), where the collision time of molecular clouds in the galactic disk is on the order of $\sim 10^{7}$ y. 
On the other hand, the collision time of a cloud on the bow head of MBS by galactic shock wave from the up-stream side and by HII expansion from the down-stream side is about $\sim 10^6$ y, an order of magnitude shorter (Sofue 2018), which is thanks to the faster motion of the gas across the shocks.  

We may call this scenario the galactic sequential SF (GSSF), which may be summarized as follows. The interstellar gas flowing from the up-stream side of the arm by the galactic rotation is compressed by the galactic shock wave. Not only molecular clouds and GMC, but also HI gas are shocked to form a broad bow structure, where HI-to-\Htwo transition takes place to form molecular gas, forming a large moleculr complex. An arched structure is formed due to the bow shock by the encounter with the expanding HII sphere produced by a preceding high mass SF. The concave-side molecular gas in the MBS is further compressed backward by the expanding HII gas from the down-stream side. This "dual-side compression" further enhances the SF in the densest region at the concave edge of the MBS, which triggers the next generation SF and MBS compression in the up-stream side. Figure \ref{fig8} illustrates the proposed scenario.

 \subsection{Comparison with other SF mechanisms}

The GSSF may work complementarily to the currently accepted mechanisms such as the sequential SF and cloud collision mechanisms, or it rather integrates these mechanisms in galactic scale. In fact, Torii et al. (2018) argued for cloud collisions in the densest region of the MBS G24.4, where a dense molecular cloud is locally associated with HII regions in the deep "bay" at G24.5+0.23 on the eastern MBS edge, and another HII region at the eastern end of the molecular "peninsula" at G24.4+0.07. It is interesting to point out that the three HII regions studied by Torii et al. are coherently located at the eastern edges of the moleuclar clouds, which indicates an ordered alignment of the clouds and SF sites. Such an alignment is in favor of the GSSF scenario, where the SF occurs, surfining on the gas flow in the galactic rotation.

The advantage of GSSF would be, thus, three folds: Firstly, GSSF lasts for galactic time scale, keeping the high rate as long as the Galaxy rotates with spiral arms. Secondly, all the interstellar gases, from tiny clouds to GMC, from diffuse gas to filaments, and even HI, are equally given the chance to take part in the SF at the large-scale shocks. Thirdly, GSSF integrates the individual cloud-size SF mechanisms to a systematic and efficient SF mechanism working on spiral arm scales. 

The GSSF may also solve some problems in the current SF theories. The HI gas may be now considered to play more direct role in the SF through rapid transition to \Htwo in the MBS. The fade-out SF problem for a single or collided clouds after exhaustion by SF can be solved by the continuously inflowing clouds and compressed gases. The too small collision probability between the smallest-mass clouds with the smallest cross sections, which share the major fraction of molecular mass according to the cloud mass function, significantly increases to unity, so that any clouds are given equal chance with probability 100\% to participate in the SF.
 
\subsection{Relation to Schmidt law}

\begin{figure} 
\begin{center}         
\includegraphics[width=6cm]{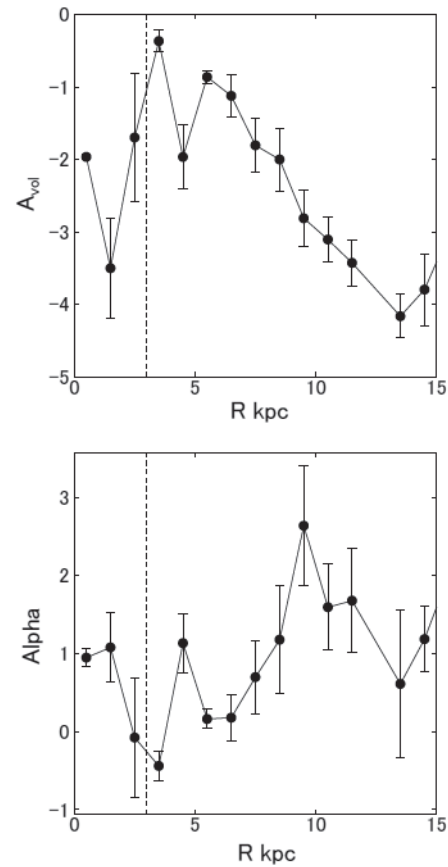}       
\end{center}
\caption{Radial variations of volume SFR coefficient $A_{\rm vol}=\log\ A$ and power-law index $\alpha$ of the Schmidt law in the Milky Way as reproduced from Sofue (2017b). Note that $A_{\rm vol}$ attains maximum at $R\sim 3$ kpc (dashed line), and $\alpha$ is minimum indicating weakest dependence of SFR on gas density there.  }
\label{fig11}
\end{figure}

The chance for any components of ISM to form stars by GSSF comes every half a galactic rotation, $t\sim \pi R/(\Vrot-\Vpat)\sim 7\times 10^7$ y at $R= 3$ kpc. The Schmidt law analysis in the Milky Way (Sofue 2017b) showed that the volume SF efficiency attains its maximum at $R\sim 3$ kpc, as shown in figure \ref{fig11}, where is plotted the coefficient of volume SFR against the galacto-centric distance $R$ of the Schmidt-Kennicut law,
$A_{\rm vol}(R)={\rm log}\ A(R)$, and power-law index, $\alpha(R)$, 
defined through
\be 
{SFR(R)\/[\Msun {\rm kpc^{-3} Gy^{-1}}]}={A(R)\/[\Msun {\rm kpc^{-3} Gy^{-1}}]} \({\rho_{\rm gas} \/[{\rm H\ cm^{-3}}]}\)^{\alpha(R)},
\ee
where $\rho_{\rm gas}$ is the volume gas density including both \Htwo and HI gases. The coefficent $A(R)$ represents specific SFR, when the gas density is fixed to a constant value, e.g. 1 H cm$^{-3}$, or the SF efficiency.
The fact that $A(R)$ attains the maximum significantly inside the density maximum at the 4-kpc molecular ring implies that the SFR is related to the arm-passage frequency, such that the innermost arm has the highest SFR.  

Another interesting feature is the mild power index of the Schmidt law inside the 4-kpc molecular ring, where $\alpha \sim 0.5$, and mildest at $\sim 3$ kpc with $\alpha \le \sim 0$. The milder slope indicates weaker dependence of SFR on the gas density, which is in favor of the GSSF's equal-chance participation of all kinds of ISM.  

\subsection{Rayleigh-Taylor instability  and elephant trunks }

At the interface between MBS and HII region, the shock-compressed inner molecular edge of MBS suffers from further dynamical effects. A well known effect is elephant trunks (ET) due to the Rayleigh-Taylor instability (RTI) caused by the acceleration of dense MBS gas by low-density HII gas. 

Elephant-trunk features are indeed observed in the CO channel maps near the MBS edge, particularly in the close up map in figure \ref{fig6}, where the CO edge is wavy with wavelength (distance between tongues) of about $0\deg.6$ (20 pc) and full amplitude (length of tongue) of $0\deg.1$ (13 pc). Top of each tongue (maximum amplitude position) has a high-density cloud. More developed giant ETs have been found in G24 and G31 regions (Sofue 2019).

It is interesting to note that the born HII regions are located on the higher-longitude sides of the tongues, and some isolated tongues are tailing toward lower longitudes. Such a coherent alignment of the molecular tongues and HII regions would be an evidence that the SF in this area occurred by the backward-compression of molecular gas by the feedback from preceding SF activity. 

\subsection{Spiral arm or expanding ring?}

The origin of the non-circular motion revealed in the LV diagram in figure \ref{fig1} is still under discussion. One idea is that it is due to a spiral arm with large orbital deformation by the bar potential as illustrated by a dashed ellipse in figure \ref{fig1} (e.g., Binney 1991). In this case the gas is compressed by the acceleration in the bar potential, and the galactic shock theory applies, on which is based this paper.

Another idea is expanding ring driven by an explosive event at the Galactic Center as indicated by the dashed circle in figure \ref{fig1} (e.g., Oort 1977 for review). In this case, the gas is compressed at the expanding front of the ring, and the basic shock structure and kinematics before star formation at the tangent point is same, and the present model also applies.

\section{Summary}

We found a new molecular bow shock (MBS) at G24.4+00+112 $(l\sim 24\deg.4,\ b\sim 0\deg, \ \vlsr \sim 112$ \kms) using the FUGIN \co-line survey taken with the Nobeyama 45-m telescope at $20''$ (0.71 pc) resolution. The terminal velocity at 112 \kms made it possible to uniquely determine the distance and location of the MBS  at the tangent point of the 3-kpc expanding arm (Norma arm) at 7.3 kpc. The MBS extends over 160 pc perpendicularly to the galactic plane and the mass is estimated to be $\sim 1.6 \times 10^6 \Msun$. 

The MBS is embedded in a larger-scale HI bow, more extended in the western (up-stream) side. The molecular fraction increases from up- to down-stream sides, attaining the maximum at the MBS edge, where the fraction is saturated at $\fmol \sim 95$ \%. This may be the evidence for the transition of HI to \Htwo in the spiral arm due to compression at the galactic shock wave.

Comparison with radio continuum and IR maps showed that the MBS is concave to the HII shell G24.6+00 and the ring of compact HII regions, which are at the same distance from the recombination line velocities. The inner edge of the MBS is extremely sharp, indicating that the molecular gas is compressed by a shock wave due to the expansion of the HII shell from east to west. This compression direction is inverse to that expected from the galactic shock, which shows that the MBS is shock-compressed from both sides: by the galactic shock from west, and by HII expansion from the east. We argued that the dual compression triggers SF at much higher rate than by the cloud collision mechanism.

The on-going SF is evidenced by the existence of compact HII regions along the edge of HII shell, which is associated with giant molecular elephant trunks (tongues) formed by the Raighly-Taylor instability at the HII-MBS interface. 
Based on the data and considerations, we proposed a scenario of galactic sequential SF (GSSF) due to dual-side compression by the galactic shock and backward shock of HII gas expansion as the feedback from preceding SF activity (figure \ref{fig8}).

\vskip 5mm
{\bf Aknowledgements}
The author expresses his sincere thanks to the authors of the survey data used in the analyses, particularly to Prof. T. Umemoto of NAOJ for the FUGIN CO data. The HI data were taken from the THOR survey. Radio continuum data were taken from the VGPS survey. IR data were taken from the GLIMPSE/ATLASGAL data archives. Data analysis was carried out at the Astronomy Data Center of the National Astronomical Observatory of Japan. 

\begin{appendix}
\section{FUGIN chanel maps in the \co line}

In this appendix, we show channel maps of the FUGIN \co line data around the G24 region. 

\begin{figure}
\begin{center}       
\includegraphics[width=6.5cm]{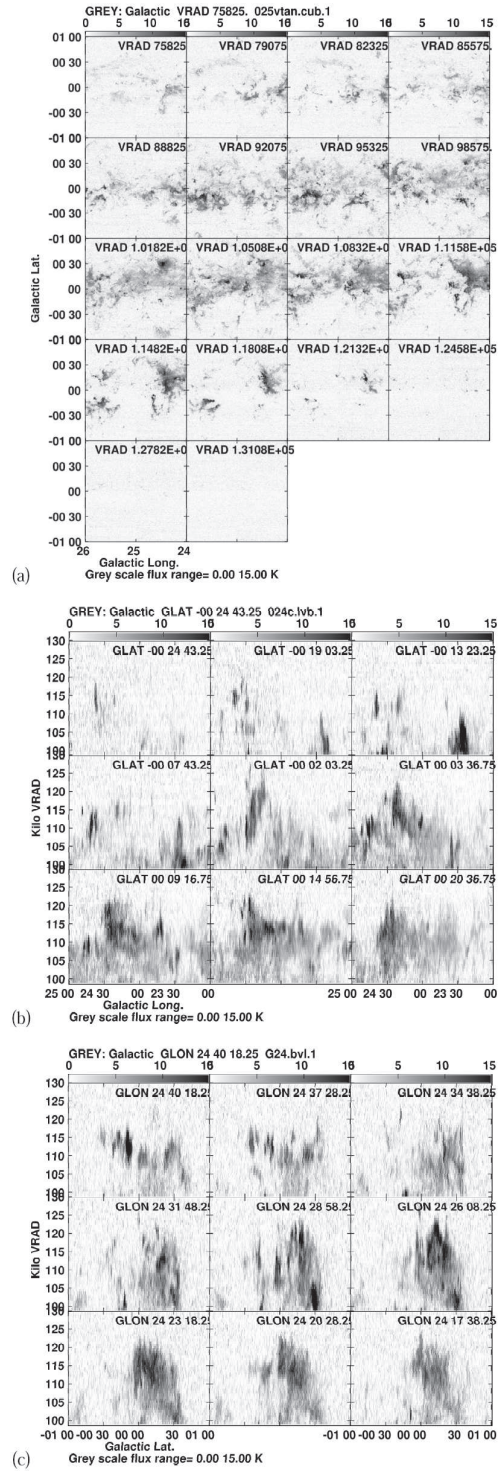}      
\caption{(a) \co $\Tb$ channel maps. The MBS appears as the bright arc at G24.4 with the peak intensity at 112 \kms. 
(b) Longitude-velocity diagrams near the tangent velocity at various latitudes. (c) Latitude-velocity maps across the MBS at different longitudes.}
\end{center}
\label{figA}  
\end{figure}

Figure \ref{figA}(a) shows channel maps of the brightness temperature around the tengent velocity at $\sim$G24. 

Figures \ref{figA}(b) and (c) show longitude-velocity diagrams at different latitudes, and latitude-velocity diagrams at different longitudes, respectively.

\end{appendix}

\end{document}